# DYNAMIC INTERSECTORAL MODELS WITH POWER-LAW MEMORY

**Valentina V. Tarasova**,
Lomonosov Moscow State University Business School, Lomonosov Moscow State University,
Moscow 119991, Russia; E-mail: v.v.tarasova@mail.ru;
**Vasily E. Tarasov**,
Skobeltsyn Institute of Nuclear Physics, Lomonosov Moscow State University,
Moscow 119991, Russia; E-mail: tarasov@theory.sinp.msu.ru

**Abstract.** Intersectoral dynamic models with power-law memory are proposed. The equations of open and closed intersectoral models, in which the memory effects are described by the Caputo derivatives of non-integer orders, are derived. We suggest solutions of these equations, which have the form of linear combinations of the Mittag-Leffler functions and which are characterized by different effective growth rates. Examples of intersectoral dynamics with power-law memory are suggested for two sectoral cases. We formulate two principles of intersectoral dynamics with memory: the principle of changing of technological growth rates and the principle of domination change. It has been shown that in the input–output economic dynamics the effects of fading memory can change the economic growth rate and dominant behavior of economic sectors.

## 1. INTRODUCTION

Dynamic models describe the dynamics of intersectoral balance of the gross product and the final product (national income) in the economy. These models are based on the equations of input-output balance in monetary terms that describe the production and distribution of the gross and final products between sectors. It takes into account the intersectoral production links, the use of material resources, the production and distribution of the national income. In the balance equation each sector is considered twice as a consumer and as a producer. This leads us to matrix form of the balance equations. A distinguishing feature of the intersectoral model is a description of the "input-output" balance equations in the matrix form. The matrix equation of intersectoral balance assumes that each product has only one production (one sector), and each production (industry) produces only one type of product. In the dynamic intersectoral models, the exogenous and endogenous variables are described by matrices.

One of the most famous models is the dynamic intersectoral model developed by Nobel laureate Wassily W. Leontief. The dynamic intersectoral model has been proposed by Leontief in the fifties of the XX century [1, 2, 3, 4]. The Leontief dynamic model is an economic model of growth of gross national product and national income.

The dynamic intersectoral models use different assumptions and are not taken into account some of the economic factors. One of the assumptions, which are commonly used in dynamic intersectoral models, is the neglect of the memory effects. In fact, the dynamic intersectoral models assumed that economic agents cannot remember history of changes of the endogenous and exogenous variables. As a result, we can say that these models describe only processes, in which all agents have full amnesia. The memory effects can play an important role in economics and natural sciences [5, 6, 7, 8, 9].

In macroeconomics, the memory can be considered as a property of economic processes, which characterizes the dependence of this process at a given time on the states in the past. In economic process with memory the endogenous and exogenous variables at a given time depend on their values at previous instants of time. In macroeconomic processes, the behavior of economic agents based not only on information on the state of the process $\{t, X(t)\}$ at a given moment of time t, but also on the use of information about the states $\{\tau, X(\tau)\}$ at time instants $\tau \in [0, t]$. A presence of memory in the processes means that there is an endogenous variable that depends not only on values of an exogenous variable at present time, but also on its values at previous instants of time. A memory effect is related with the fact that the same change of the exogenous variable can leads to the different change of the corresponding endogenous variable. This leads us to the multivalued dependencies of these variables. The multivalued dependencies are caused by the fact that the economic agents remember previous changes of these variables, and therefore can react differently. As a result, an identical change of the exogenous variable may lead to the different dynamics of endogenous variables.

In economics, the concept of memory can be considered by analogy with fractional dynamics [8, p. 394-396]. In this paper, we propose a method of accounting the power-law memory in the construction of dynamic intersectoral models in the form of generalization of the dynamic Leontief model with continuous time. As a mathematical tool we use the theory of differential equations with derivatives of non-integer order [10, 11, 12, 13, 14]. Our consideration is based on the concept of the accelerator with memory and marginal value of non-integer order, which are suggested in [15, 16, 17, 18, 19]. This paper presents and analyzes the solutions of the equations of the closed and open intersectoral dynamic models by using the solutions on fractional differential equations, which are considered in [12, 13, 20, 21, 22].

## 2. DYNAMIC INTERSECTORAL MODEL WITHOUT MEMORY

Let us consider intersectoral model, where we assume that n kinds of products are produced and used. Each sector produces only one type of products, and each product is produced in a certain sector. Characteristics of production processes are assumed to be constant, that is, we will not take into account that technological progress can lead to changes in production technology. In addition, the import of goods and materials and the use of non-renewable resources will not be considered in the model. Dynamic intersectoral model will be formulated in the framework of continuous time approach.

Let us describe a derivation of input-output balance of equation of dynamic Leontief model. In the Leontief model, the gross product (gross output) is described by the vector $X(t) = (X_k(t))$, and it is divided into two parts

$$X(t) = Z(t) + Y(t), \tag{1}$$

where $Y(t) = (Y_k(t))$ is a vector of the final product; $Z(t) = (Z_k(t))$ is the vector of the intermediate product, where k=1,…,n are production sectors. The final product is distributed to the investments and the non-productive consumption

$$Y(t) = I(t) + C(t), \qquad (2)$$

where $I(t) = (I_k(t))$ is a vector of investments; $C(t) = (C_k(t))$ is the vector of products of nonproductive consumption (including non-productive accumulation), where k = 1,…, n are production sectors. The Leontief dynamic model assumes the performance of the balance equations (1) and (2) for any t>0. These equations describe the dynamic equilibrium of the economy as a whole. For this reason, the dynamic Leontief model is a dynamic model of the "input-output" balance. Substituting the expression of the final product (2) into formula (1), we obtain the balance equation

$$X(t) = Z(t) + I(t) + C(t). \qquad (3)$$

To get the Leontief model equation from the balance equation (3), it is necessary to eliminate the endogenous (internal) variables Z(t) and I(t). To do this, we should give dependence of Z(t) and I(t) on the exogenous variable X(t).

The Leontief model assumes constancy of coefficients of direct material costs of production. The dependence of the intermediate product on the gross product is assumed in the form of direct proportionality. This allows us to express the vector of intermediate products Z(t) through the multiplication of the matrix of direct material costs A and the vector of gross product X(t) in the form of the matrix multiplier equation

$$Z(t) = A \cdot X(t), \qquad (4)$$

where $A = (a_{ij})$ is the square matrix of n-th order with coefficients $a_{ij}$, which describe the direct material costs of i-th sector (i=1,…,n) in the production of a unit of output j-th sector (j=1,…,n). The matrix A is assumed to be constant, that is, it does not change with time. In the dynamic Leontief model the coefficients $a_{ij}$ include not only the direct material costs, but also the costs of disposal compensation and repair the basic production assets. Therefore, the elements of the main diagonal of the matrix A are nonzero.

The dynamic Leontief model is based on the assumption of the relationship between the accumulation and of the growth of gross output. This relationship is implemented by using matrix of capital intensity of production growth. In addition, it is assumed instantaneous transformation of investment in a gain of fixed assets and instantaneous of return of these funds in the production (in the outputs), that is, the model disregards delay. Time is supposed continuous. This allows us to use the theory of differential equations. Dependence of the vector I(t) of capital investments on the vector X(t) of the gross product is described by the matrix accelerator equation

$$I(t) = B \cdot \frac{dX(t)}{dt}, \qquad (5)$$

where $I(t) = (I_k(t))$ is the vector of investments, $B = (b_{ij})$ is square matrix of the n-th order of the coefficients $b_{ij}$ of an incremental capital intensity of production. Coefficients $b_{ij}$ describe the expenses of production of i-th sector for increase in production in j-the sector. Matrix B is assumed non-degenerate, i.e., the determinant of the matrix B is different from zero. For the square matrix B, this condition is equivalent to the reversibility of the matrix, i.e. the existence of an inverse matrix $B^{-1}$. In general, the matrix B can have zero line, for example, for sectors, which produce only consumer goods. For the reversibility of the matrix B of intersectoral dynamic models, we can consider only the equation for sectors that form the fixed assets.

Substitution of expressions (4) and (5) into equation (3), we get the equation of the dynamic Leontief model for gross production (gross output) in the form of the matrix differential equation

$$B \cdot \frac{dX(t)}{dt} + (A-E) \cdot X(t) + C(t) = 0, \qquad (6)$$

where E is the unit diagonal matrix of n-th order. The economic sense is made by only such solutions of equation (6), for which $X(t) \geq 0$.

The equation for the final product (national income), which is described by the vector Y(t), can be obtained from equation (6). To do this, we substitute the expression of Z(t), which is given by formula (4), into equation (1) and then we express Y(t). As a result, we obtain the expression

$$Y(t) = (E-A) \cdot X(t), \qquad (7)$$

Equation (7) allows us to find the vector Y(t) of final product if the vector X(t) of the gross product is given. Substituting (7), which is written in the form $X(t) = (E-A)^{-1} \cdot Y(t)$, into equation (6) and taking into account the constancy of the matrix A, we get equation of the final product (national income) in the form

$$B \cdot (E-A)^{-1} \cdot \frac{dY(t)}{dt} - Y(t) + C(t) = 0, \qquad (8)$$

where $B \cdot (E-A)^{-1}$ is the matrix of the full incremental capital intensity. Note that the economic models without the foreign trade have an economic meaning only if solutions of equation (8) satisfy the condition $Y(t) \geq 0$.

Let us note the some features of the dynamic intersectoral model.

(a) The coefficients of direct material costs $a_{ij}$ and the coefficients incremental capital $b_{ij}$ are assumed to be constant, i.e. the matrix A and B are constant in time. In the general case this is not true, especially when we consider the long time intervals.

(b) Increase of production is instantaneous, when we change the investments. This means that we ignore the effects of the time delay between the final product and to the investment.

(c) Investments I(t) at time t determined by the change in the gross product X(t) only in an infinitesimal neighborhood of the point in time, i.e. nearest infinitesimal past. This implies neglect of memory effects. In other words, in fact it is supposed to instant amnesia for all economic agents, i.e. the instant forgetting of the history of changes of the gross product and investments.

Intersectoral model is said to be closed if we assume the absence of non-productive consumption, i.e. C(t)=0. The general solution of the system of differential equations (8) with C(t) = 0 can be written [8, 9] in the form

$$Y(t) = \sum_{k=1}^{n} c_k \cdot Y_k \cdot \exp(\lambda_k \cdot t), \qquad (9)$$

where $\lambda_k$ are eigenvalues of the matrix $\Lambda := (E - A) \cdot B^{-1}$, the vectors $Y_k = (Y_{kj})$ are the eigenvectors, which correspond to the eigenvalues $\lambda_k$, i.e. $\Lambda \cdot Y_k = \lambda_k \cdot Y_k$, and the coefficients $c_k$ are determined from the initial condition $\sum_{k=1}^{n} c_k \cdot Y_k = Y(0)$.

The solution (9) is a combination of exponential functions $\exp(\lambda_k \cdot t)$ with different rates of growth $\lambda_k$. Therefore, in general, the dynamics of an economic process by the path $Y(t) = Y(0) \cdot \exp(\lambda \cdot t)$ with a uniform growth rate $\lambda$ for all sectors is impossible. As a result, economic growth will be with permanent structural changes. This fact significantly distinguishes the intersectoral model from the macroeconomic models such as the natural growth model, the Harrod-Domar model, the Keynes model and other. However, there is some correlation between the solution of the dynamic intersectoral models and the solutions of macroeconomic growth models. This relationship is based on the existence of an eigenvalue $s_{max}$ with the highest absolute value for the matrix $S = B \cdot (E - A)^{-1} = \Lambda^{-1}$ of the full incremental capital intensity. The existence of such eigenvalue follows from the Perron's theorem and its generalization by Frobenius. The Perron's theorem asserts [23] that for the positive matrix always exists a unique eigenvalue $s_{max}$ with the highest absolute value. The number $s_{max}$ is positive, and it is a simple root of the characteristic equation. The corresponding eigenvector can be chosen positive. The Frobenius theorem asserts [24, p. 354-355] that an irreducible non-negative matrix always has a positive eigenvalue $s_{max}$, which is a simple root of the characteristic equation. All other eigenvalue do not exceed modulo the number $s_{max}$, and this number corresponds to the eigenvector with positive coordinates. The number $s_{max}$ of the matrix S is often called the Frobenius-Perron number.

In the intersectoral dynamic models the eigenvalue $\lambda_s = 1/s_{max}$ of the matrix $\Lambda = S^{-1}$, is called the technological growth rate [8, p. 126]. It is important to note that the eigenvectors $Y_k$, which corresponds to the eigenvalues $\lambda_k$ that differ from $\lambda_s$ ($\lambda_k \neq \lambda_s$), necessarily have the coordinates with different signs.

Solutions (9) of the closed intersectoral model are a combination of exponential functions $\exp(\lambda_k \cdot t)$ with different rates of growth $\lambda_k$. In the solution the term, which has the maximum real part $\lambda_k$ and corresponds to $c_k \neq 0$, begins to dominate at $t \to \infty$. If the dominant is the term with $\lambda_s = 1/s_{max}$, then the rate of growth of final products tends to the technological growth rate at $t \to \infty$ in all sectors. In this case, the sectoral structure of the national income in the limit ($t \to \infty$) will be determined by the proportions of the coordinates of the corresponding eigenvector $Y_k$.

If the dominant term will have the growth rate $\lambda_k \neq \lambda_s = 1/s_{max}$, then the dynamics of $Y(t)$ at $t \to \infty$ will be determined by the corresponding eigenvector $Y_k$, the coordinates of which have different signs. Therefore this solutions $Y(t)$ at $t \to \infty$ will necessarily have the negative components, which means that the solutions lose the economic sense [8, p. 127]. As a result, solutions (9) are economically unacceptable if the terms with the growth rate $\lambda_k$, which different from $\lambda_s = 1/s_{max}$, are dominated. This feature of closed intersectoral models distinguishes this model from its macroeconomic analogues, in which the solutions are inadmissible because of the overestimated requirements to the growth of the consumption.

### 3. DYNAMIC INTERSECTORAL MODEL WITH POWER-LAW MEMORY

Equations (6) and (8) describe the dynamic intersectoral models, assume that the relationship between the vector I(t) of capital and the vector X(t) of the gross product is given by linear accelerator equation (5). Equation (5) contains derivative of the first order. The use of this derivative means an instantaneous change of the endogenous variable I(t) at change of the exogenous variable X(t). In other words, the investment at time t is determined by the properties of the vector of the gross product in infinitely small neighborhood of this point in time. Therefore accelerator equation (5) does not take into account the memory effects. As a result, the differential equations (6) and (8) in fact can be used only to describe the economy, in which all economic agents have an instant amnesia. This restriction significantly reduces the scope of the intersectoral models to describe the real economic processes. In many cases, economic agents may remember the story of changes of the gross product and previous investments. Therefore, the neglect of the memory effects can lead to incorrect descriptions of economic processes.

To take into account the power-law memory we can apply the mathematical formalism of derivatives and integrals of non-integer orders [10, 11, 12, 13, 14]. The concepts of the marginal values of non-integer order and the accelerator with memory have been proposed in [15, 16, 17, 18, 19]. The use the concept of the accelerator with memory [15, 16] allows us to generalize the dynamic Leontief model for economic processes with memory of power-law fading.

To describe the dynamic power-law memory in intersectoral models, we should use a generalization of equation (5), which describes the relationship between the vector of investment I(t) and the vector X(t) of the gross product. The concept of the marginal value of non-integer order, which is proposed in [17, 18, 19], allows us to get the equation of the accelerator with memory [15, 16]. The linear equation of accelerator with power-law memory has the form

$$I(t) = B \cdot (D_{0+}^{\alpha} X)(t). \tag{10}$$

An explanation of the suggested form of the accelerator equation, which takes into account the power-law memory, is given in the Appendix of this paper. Here $(D_{0+}^{\alpha} X)(t)$ is the Caputo fractional derivative of the order $\alpha \geq 0$ [11, 12], which is defined by the equation

$$(D_{0+}^{\alpha} X)(t) := \frac{1}{\Gamma(n-\alpha)} \int_0^t \frac{X^{(n)}(\tau) d\tau}{(t-\tau)^{\alpha-n+1}}, \tag{11}$$

where $\Gamma(\alpha)$ is the gamma function, $X^{(n)}(\tau)$ is the derivative of the integer order $n:=[\alpha]+1$ of the function $X(\tau)$ with respect to time variable $\tau$: $0<\tau<t$. Equation (11) assumes that the function $X(\tau)$ has the derivatives of integer order up to the (n-1)th order, which are absolutely continuous functions on the interval [0,t]. For $\alpha=1$ equation (10) gives equation (5). In order to have easily interpretable dimension of the quantities, we can use the time t as a dimensionless variable.

For intersectoral dynamic model, equation (10) actually assumes a uniform parameter of memory fading for all sectors of economy. In this paper, we first consider the dynamic models with a uniform fading parameter $0<\alpha<2$. Then we will generalize the model to the case of sectoral memory, which is determined by the vector $\alpha = (\alpha_1, \ldots, \alpha_n)$ of fading order, where $\alpha_k$ is the parameter of the memory fading in the k-th sector of economy.

Substitution of equations (10) and (4) into balance equation (3), we get the fractional differential equation

$$B \cdot (D_{0+}^{\alpha} X)(t) + (A-E) \cdot X(t) + C(t) = 0. \tag{12}$$

Equation (12) describes dynamic intersectoral models, which take into account power-law memory and generalize Leontief model that is described by equation (6). For $\alpha=1$ equation (12) has the form (6).

The equation of the final product (national income) can be obtained from equation (12) by using expression (7) in the form $X(t) = (E-A)^{-1} \cdot Y(t)$. Since the matrix A is assumed constant, then substituting the expression $X(t) = (E-A)^{-1} \cdot Y(t)$ into equation (12), we obtain

$$B \cdot (E-A)^{-1} \cdot (D_{0+}^{\alpha} Y)(t) - Y(t) + C(t) = 0, \tag{13}$$

Equation (13) describes dynamics of final product Y(t) in the dynamic intersectoral model with power-law memory. The matrix $B \cdot (E-A)^{-1}$ is called the matrix of the full incremental capital intensity.

Let us consider closed dynamic model for the gross and final products. The closed models are described by equations (12) and (13) with zero non-productive consumption (C(t)=0). Using (12), the matrix equation for the gross product of X(t) at C(t)=0 has the form of the fractional differential equation

$$B \cdot (D_{0+}^{\alpha} X)(t) + (A-E) \cdot X(t) = 0. \tag{14}$$

For $\alpha=1$ equation (14) takes the form

$$B \cdot \frac{dX(t)}{dt} + (A-E) \cdot X(t) = 0. \tag{15}$$

Using (13) with C(t) = 0, we get the equation for the final product Y(t) in the form
$$Y(t) = B \cdot (E - A)^{-1} \cdot (D_{0+}^\alpha Y)(t), \qquad (16)$$
where $B \cdot (E - A)^{-1}$ is the matrix of the full incremental capital intensity. Equations (14) and (16) describe dynamics of sectoral structure of the gross and final products in the closed dynamic intersectoral model with power-law memory. Solutions of closed model equations (14) and (16) characterize the extreme (limit) technological possibilities of production sectors, for given A and B, when the entire national income is directed to the expanded reproduction.

Equations (14) and (16) are the systems of linear fractional differential equations. If the matrix of the full incremental capital intensity is reversible, then equations (14) and (16) can be written as
$$(D_{0+}^\alpha X)(t) = B^{-1} \cdot (E - A) \cdot X(t), \qquad (17)$$
$$(D_{0+}^\alpha Y)(t) = (E - A) \cdot B^{-1} \cdot Y(t). \qquad (18)$$
Equations (17) and (18) describe a closed intersectoral dynamic model with the power-law memory, which has a uniform parameter for all sectors of economy. It is well-known [12, p. 142] and [20] that the solution of the fractional differential equation
$$(D_{0+}^\alpha Y)(t) = \Lambda \cdot Y(t) \qquad (19)$$
with the Caputo fractional derivative (11) of the order 0<α<1, and the matrix $\Lambda = (E - A) \cdot B^{-1}$, can be represented by the expression
$$Y(t) = E_\alpha[\Lambda \cdot t^\alpha] \cdot Y(0), \qquad (20)$$
where $E_\alpha[\Lambda \cdot t^\alpha]$ is the matrix Mittag-Leffler function [13, p. 142] that is defined by the formula
$$E_\alpha[\Lambda \cdot t^\alpha] := \sum_{k=0}^\infty \frac{t^{\alpha \cdot k}}{\Gamma(\alpha k + 1)} \cdot \Lambda^k. \qquad (21)$$
The general solution of fractional differential equations (18) can be written as
$$Y(t) = \sum_{k=1}^n c_k \cdot Y_k \cdot E_\alpha[\lambda_k \cdot t^\alpha], \qquad (22)$$
where $\lambda_k$ is the eigenvalues of the matrix $\Lambda := (E - A) \cdot B^{-1}$, and $Y_k = (Y_{kj})$ are the corresponding eigenvectors such that
$$\Lambda \cdot Y_k = \lambda_k \cdot Y_k. \qquad (23)$$
The coefficients $c_k$ are determined from the initial conditions by the equation
$$\sum_{k=1}^n c_k \cdot Y_k = Y(0). \qquad (24)$$
Using $E_1[z] = \exp(z)$, we get that solution (22) with α = 1 takes the form
$$Y(t) = \sum_{k=1}^n c_k \cdot Y_k \cdot \exp(\lambda_k \cdot t). \qquad (25)$$
Thus, solution (22) contains solution (25) of the standard closed intersectoral model as a special case that corresponds to α=1.

Similarly, the general solution of equation (17) can be written in the form
$$X(t) = \sum_{k=1}^n d_k \cdot X_k \cdot E_\alpha[\omega_k \cdot t^\alpha], \qquad (26)$$
where $\omega_k$ are the eigenvalues of the matrix $\Omega := B^{-1} \cdot (E - A)$; the vectors $X_k = (X_{ki})$ are the eigenvectors of the matrix $\Omega$; and the coefficients $d_k$ are determined by the initial condition
$$\sum_{k=1}^n d_k \cdot X_k = X(0). \qquad (27)$$
Using expression (7) and solution (25), the solution of equation (17) can be written as
$$X(t) = \sum_{k=1}^n c_k \cdot (E - A)^{-1} \cdot Y_k \cdot E_\alpha[\lambda_k \cdot t^\alpha], \qquad (28)$$
which is equivalent to expression (26), where $\lambda_k = \omega_k$.

The solution is a linear combination of the Mittag-Leffler functions $E_\alpha[\lambda_k \cdot t^\alpha]$ with different eigenvalues $\lambda_k$. Therefore, in general, the dynamics of the economic process by the trajectory $Y(t) = Y(0) \cdot E_\alpha[\lambda \cdot t^\alpha]$, i.e. with the uniform parameter λ for all sectors, is impossible. It distinguishes the intersectoral model with memory and the macroeconomic models with memory, which are considered in [27, 28, 29, 30, 31, 32, 33, 34]. As a result, economic dynamics will be realized with structural changes.

Note that there is a relationship between solutions (25), (26), (28) of the dynamic intersectoral model with memory and the solutions of the macroeconomic models with memory [27, 28, 29, 30, 31, 32, 33, 34]. This relationship is based, as well as for the standard model, on the existence of the eigenvalue $s_{max}$ with the highest absolute value for the matrix $S = B \cdot (E - A)^{-1} = \Lambda^{-1}$ of the full incremental capital intensity. At the same time all the other modules of the eigenvalues of the matrix S does not exceed the number of $s_{max}$, which is the Frobenius-Perron number. The corresponding eigenvector may be selected so that all its coordinates will be positive. In this case the eigenvectors $Y_k$, which correspond to the eigenvalues $\lambda_k \neq \lambda_s$, have the coordinates of different signs. The existence of such eigenvalue follows from the Perron's theorem [14, p. 319] and the Frobenius theorem [15, p. 354-355]. Some roots $\lambda_k$ of the characteristic equation of the matrix $\Lambda := (E - A) \cdot B^{-1}$ can be complex. Thus each complex root $\lambda_k = \text{Re}(\lambda_k) + i \cdot \text{Im}(\lambda_k)$ corresponds to the conjugate root $\bar{\lambda}_k = \text{Re}(\lambda_k) - i \cdot \text{Im}(\lambda_k)$. Each pair of complex conjugate roots will be presented by a pair of terms in the solution. In this case, we have oscillations with the constant rate, which is equal to $\text{Im}(\lambda_k)$, and the variable amplitude.

Let us now analyze the dominant behavior at $t \to \infty$ for intersectoral dynamic model with memory. For this we use the asymptotic formula of the Mittag-Leffler functions $E_\alpha[\lambda_k \cdot t^\alpha]$ at $t \to \infty$. Using formulas (3.4.14) and (3.4.15) of [25, p. 25-26], we obtain for 0<α<2 and $t \to \infty$ the equation

$$E_\alpha[\lambda \cdot t^\alpha] = \frac{1}{\alpha} \cdot \exp(\lambda^{1/\alpha} \cdot t) - \sum_{k=1}^{m} \frac{\lambda^{-k}}{\Gamma(1-\alpha \cdot k)} \cdot \frac{1}{t^{\alpha \cdot k}} + O\left(\frac{1}{t^{\alpha \cdot (m+1)}}\right) \qquad (29)$$

for real values of λ and for complex roots with $|\arg(\lambda)| \leq \theta$, where

$$\arg(\lambda) := \text{arctg}\left(\frac{\text{Im}(\lambda)}{\text{Re}(\lambda)}\right), \frac{\pi\alpha}{2} < \theta < \min\{\pi, \pi\alpha\}. \qquad (30)$$

For complex values of λ, for which $\theta \leq |\arg(\lambda)| \leq \pi$, we have

$$E_\alpha[\lambda \cdot t^\alpha] = -\sum_{k=1}^{m} \frac{\lambda^{-k}}{\Gamma(1-\alpha \cdot k)} \cdot \frac{1}{t^{\alpha \cdot k}} + O\left(\frac{1}{t^{\alpha \cdot (m+1)}}\right), \qquad (31)$$

where $\Gamma(1 - \alpha) < 0$ for 0<α<1 and $\Gamma(1 - \alpha) > 0$ for 1<α<2.

As a result, for the real values of λ, we find that the growth rate of the intersectoral models with memory do not coincide with the eigenvalues $\lambda_k$ of the matrix Λ. The growth rate is equal to the values

$$\lambda_{k,eff}(\alpha) := \lambda_k^{1/\alpha}, \qquad (32)$$

which will be interpreted as the effective growth rates. For complex values of λ, for which $\theta \leq |\arg(\lambda)| \leq \pi$, the dynamics at $t \to \infty$ is determined by the inverse proportionality $t^{-\alpha}$ instead of the exponential functions. In the presence of terms with the exponential behavior at $t \to \infty$, the contribution of the terms with $t^{-\alpha}$ can be neglected at $t \to \infty$.

Solution (22) of equation (18) is a combination of the Mittag-Leffler functions $E_\alpha[\lambda_k \cdot t^\alpha]$ with different parameter $\lambda_k$, which correspond to different effective growth rates $\lambda_{k,eff}(\alpha)$. In solution (22) at $t \to \infty$ will dominate the term with the maximum real part of $\lambda_{k,eff}(\alpha)$, for which $c_k \neq 0$ and $|\arg(\lambda)| \leq \theta$, where θ satisfies inequality (30). If the dominant term has the effective growth rate

$$\lambda_{s,eff}(\alpha) = (1/s_{max})^{1/\alpha}, \qquad (33)$$

where $s_{max}$ is the Frobenius-Perron number of the matrix S, then the growth rate of final products in all sectors at $t \to \infty$ tends to the effective technological growth rate (33). In this case, the sectoral structure of the national income at $t \to \infty$ is determined by the proportion between the coordinates of the eigenvector $Y_s$, which corresponds to $\lambda_s = 1/s_{max}$. If the dominant is a term with $\lambda_{k,eff}(\alpha) \neq \lambda_{s,eff}(\alpha)$, then the dynamics of $Y(t)$ at $t \to \infty$ is defined by the corresponding eigenvector $Y_k$, which coordinates have different signs. Therefore this solution $Y(t)$ at $t \to \infty$ will be necessarily to have negative components. This means that the solution loses an economic sense. As a result, the solutions (22), (26), (28), where the terms with $\lambda_{k,eff}(\alpha) \neq \lambda_{s,eff}(\alpha)$ are dominated, are economically unacceptable and these solutions do not have an economic sense. Note that this feature of the solution of closed models distinguishes the dynamic intersectoral model with memory from the macroeconomic models with memory [27, 28, 29, 30, 31, 32, 33, 34], in which the solution becomes unacceptable because of the large requirements for the growth of consumption.

A comparison of the growth rates of the intersectoral models with power-law memory and the standard model without memory, is given in the following table for the real values of $\lambda_k$ and $0 < \alpha < 2$.

|  | $0 < \alpha < 1$ | $1 < \alpha < 2$ |
| --- | --- | --- |
| $0 < \lambda_k < 1$ | $\lambda_{k,eff}(\alpha) < \lambda_k$ | $\lambda_{k,eff}(\alpha) > \lambda_k$ |
| $\lambda_k = 1$ | $\lambda_{k,eff}(\alpha) = \lambda_k$ | $\lambda_{k,eff}(\alpha) = \lambda_k$ |
| $\lambda_k > 1$ | $\lambda_{k,eff}(\alpha) > \lambda_k$ | $\lambda_{k,eff}(\alpha) < \lambda_k$ |

The table shows that the accounting of memory effects can significantly change the growth rates of the dynamic intersectoral models. At the same time the growth rates can increase and decrease in comparison with the standard model. If the Frobenius-Perron number of the matrix of the full incremental capital intensity is less than one ($0 < s_{max} < 1$), the corresponding growth rate can be much larger if we take into account the memory effects. For example, when $\alpha=0.2$ and $s_{max} = 0.5$, the growth rate in the model with memory is $\lambda_{s,eff}(0.5) = 32$ instead of $\lambda_s = 2$ for the standard model, i.e. it will be 16 times more.

These results allow us to formulate the following principle.

**Principle of changing of technological growth rates.** *For low rates of the technological growth of the standard intersectoral model, the accounting the memory effects with fading parameter $0<\alpha<1$ leads to a decrease in the rate of economic growth, and it leads to an increase in the growth rate for $1<\alpha<2$. For high rates of the technological growth of the standard intersectoral model, the accounting the memory effects with the fading parameter $0<\alpha <1$ leads to an increase in economic growth, and it leads to a decrease in the growth rate for $1<\alpha<2$.*

An example, which illustrates this principle, is presented in the last section of this paper.

For $1<\alpha<2$ the general solution of the fractional differential equation
$$(D_{0+}^\alpha Y)(t) = \Lambda \cdot Y(t) \tag{34}$$
with the Caputo fractional derivative of the order $1<\alpha<2$ can be written [12, p. 232] and [25] in the form
$$Y(t) = \sum_{k=1}^{n} c_{1k} \cdot Y_k \cdot E_\alpha[\lambda_k \cdot t^\alpha] + \sum_{k=1}^{n} c_{2k} \cdot Y_k \cdot t \cdot E_{\alpha,2}[\lambda_k \cdot t^\alpha], \tag{35}$$
where $E_{\alpha,2}[S]$ is the two-parametric matrix Mittag-Leffler function [12, p. 42], which is defined by the equation
$$E_{\alpha,\beta}[S] := \sum_{k=0}^{\infty} \frac{1}{\Gamma(\alpha k+\beta)} \cdot S^k. \tag{36}$$
For β=1 the function (36) takes the form of the Mittag-Leffler function (21), such that $E_{\alpha,1}[S] = E_\alpha[S]$. The numbers $\lambda_k$ are eigenvalues of the matrix $\Lambda := (E - A) \cdot B^{-1}$, and $Y_k = (Y_{kj})$ are corresponding eigenvectors such that

$\Lambda \cdot Y_k = \lambda_k \cdot Y_k.$ (37)

The coefficients $c_{1k}$ and $c_{2k}$ are determined by the initial conditions in the form

$\sum_{k=1}^{n} c_{1k} \cdot Y_k = Y(0),$ (38)

$\sum_{k=1}^{n} c_{2k} \cdot Y_k = Y^{(1)}(0),$ (39)

where $Y(0)$ and $Y^{(1)}(0)$ are the national income and the speed of change of the national income at the initial time t=0, respectively.

Let us give the asymptotic formulas for the two-parameter Mittag-Leffler function $E_{\alpha,2}[\lambda_k \cdot t^{\alpha}]$ at $t \to \infty$. Using equation (1.8.27) and (1.8.28) from [12, p. 43], we obtain for 0<α<2 and $t \to \infty$ the formula

$E_{\alpha,2}[\lambda \cdot t^{\alpha}] = \frac{\lambda^{-1/\alpha}}{\alpha} \cdot t^{-1} \cdot \exp(\lambda^{1/\alpha} \cdot t) - \sum_{k=1}^{m} \frac{\lambda^{-k}}{\Gamma(2-\alpha \cdot k)} \cdot \frac{1}{t^{\alpha \cdot k}} + O\left(\frac{1}{t^{\alpha \cdot (m+1)}}\right)$ (40)

for real values of λ and the complex values of λ, for which $|\arg(\lambda)| \leq \theta$, where $\frac{\pi\alpha}{2} < \theta < \min\{\pi, \pi\alpha\}$. For complex values of λ, for which $\theta \leq |\arg(\lambda)| \leq \pi$, we have

$E_{\alpha,2}[\lambda \cdot t^{\alpha}] = -\sum_{k=1}^{m} \frac{\lambda^{-k}}{\Gamma(2-\alpha \cdot k)} \cdot \frac{1}{t^{\alpha \cdot k}} + O\left(\frac{1}{t^{\alpha \cdot (m+1)}}\right)$ (41)

The asymptotic formulas (40) and (41) allow us to describe the dominant behavior at $t \to \infty$ in the intersectoral dynamic model with the power-law memory, which has the fading parameter 0<α<2.

## 4. OPEN DYNAMIC INTERSECTORAL MODEL WITH MEMORY

Open dynamic intersectoral model with power-law memory is described by equations (12) and (13) with non-zero non-productive consumption ($C(t) \neq 0$). Solutions of equations (12) and (13) can be obtained by using solutions of the fractional differential equations that are proposed in [12, 13, 21, 22].

Let us consider an open dynamic model with memory for the vector of the final product (national income). Open model with power-law memory is described by the matrix fractional differential equation (13). Using the reversibility of the matrix B, equation (13) can rewritten in the form

$(D_{0+}^{\alpha} Y)(t) = (E - A) \cdot B^{-1} \cdot Y(t) - (E - A) \cdot B^{-1} \cdot C(t).$ (42)

Equation (42) is inhomogeneous fractional differential equation [11, 12], which describes the open dynamic intersectoral model.

Let us consider the economic model that is described by equation (42), which takes into account the dynamics of non-productive consumption and power-law memory. The general solution of fractional differential equations (42) can be represented as the sum of the general solution $Y_0(t)$ of the closed model, which is described by homogeneous equation (18), and a particular solution $Y_C(t)$ of the inhomogeneous equation (42), such that

$Y(t) = Y_0(t) + Y_C(t).$ (43)

Let us consider the vector-function of non-productive consumption $C(t)$ in the form

$C(t) = C(0) \cdot E_{\alpha}[r \cdot t^{\alpha}],$ (44)

where we assumed that the components of non-productive consumption grow with the uniform rate in all sectors. For r=0 we have $C(t) = C(0) = \text{const}$.

A particular solution of equation (42) with $C(t)$ in the form (44) is given by the expression

$Y_C(t) = (E - r \cdot S)^{-1} \cdot C(0) \cdot E_{\alpha}[r \cdot t^{\alpha}],$ (45)

where $S = B \cdot (E - A)^{-1}$ is the matrix of the full incremental capital intensity. For $0 < \alpha < 1$ the general solution of open dynamic intersectoral model with power-law memory can be represented in the form

$$Y(t) = \sum_{k=1}^{n} c_k \cdot Y_k \cdot E_\alpha[\lambda_k \cdot t^\alpha] + (E - r \cdot S)^{-1} \cdot C(0) \cdot E_\alpha[r \cdot t^\alpha], \qquad (46)$$

where $\lambda_k$ are eigenvalues of the matrix $\Lambda := (E - A) \cdot B^{-1}$, and $Y_k = (Y_{kj})$ are corresponding eigenvectors ($\Lambda \cdot Y_k = \lambda_k \cdot Y_k$), the coefficients $c_k$ are determined from the initial condition (for the case $0 < \alpha < 1$) by the equation

$$\sum_{k=1}^{n} c_k \cdot Y_k = Y(0) - (E - r \cdot S)^{-1} \cdot C(0). \qquad (47)$$

If $C(0)=0$, then solution (46) coincides with solution (22), and condition (47) becomes (24).

For 1<α<2 solutions of open model (42) with the power-law memory are obtained similarly, i.e. by using solution (35) and conditions (38), (39) for the closed model (18) with the memory fading 1<α<2.

To get a solution of the open model, we should find the eigenvalues $\lambda_k$ of the matrix $\Lambda$, and the corresponding eigenvectors $Y_k = (Y_{kj})$. Then we should set the growth rate r of non-productive consumption and find the values of the coefficients $c_k$, which correspond to the given growth rate r, from the system of algebraic equations (47).

Standard models without memory assume that r must not exceed the rate of technological growth, due to the fact that the productivity condition of the matrix r·S gives the inequality $r < \lambda_s$ [26] and [3, p. 130]. If $r > \lambda_s$, then Y(t) grows faster than the term that contains $\lambda_s$. The matrix r·S is unproductive and therefore the vector $Y_C(t)$ will have negative components. Since some components $Y_C(t)$ is negative, then we get negative components of the vector Y(t), which means that this solution does not make economic sense. As a result, a necessary condition for the existence of economically interpretable solutions of national income is $r < \lambda_s$. In models with memory, which are characterized by uniform parameter of memory fading for all sectors, the condition $r < \lambda_s$ does not change.

## 5. INTERSECTORAL MODEL WITH SECTORAL MEMORY

In the previous sections of the paper, we have assumed a uniform parameter of memory fading for all sectors of economy. In general, the parameters of different sectors can have different memory fading. In this case, the model can be called the intersectoral model with sectoral memory. Damping of the memory will be determined by the fading vector $\alpha = (\alpha_1, \ldots, \alpha_n)$. To describe the dynamics of sectors with different fading parameters of power-law memory, we can use the matrix differentiation of non-integer order. For example, we can use the diagonal matrix operators

$$\widehat{D_{0+}^\alpha} = \text{diag}(D_{0+}^{\alpha_1}, \ldots, D_{0+}^{\alpha_n}), \qquad (48)$$

where the element of this matrix are the operators $\widehat{D_{kl}^\alpha} = \delta_{kl} \cdot D_{0+}^{\alpha_k}$, where $\alpha_k$ is the fading parameter of the k-th sector of economy. For dynamic models with two sectors, the operator (48) can be written in the form

$$\widehat{D_{0+}^\alpha} = \begin{pmatrix} D_{0+}^{\alpha_1} & 0 \\ 0 & D_{0+}^{\alpha_2} \end{pmatrix}, \qquad (49)$$

and the action of operator (49) on the vector Y(t) of the final product is given by the expression

$$(\widehat{D_{0+}^\alpha} Y)(t) = \begin{pmatrix} (D_{0+}^{\alpha_1} Y_1)(t) & 0 \\ 0 & (D_{0+}^{\alpha_2} Y_2)(t) \end{pmatrix}. \qquad (50)$$

The closed intersectoral dynamic model with sectoral memory, which is characterized by the different fading parameters in the different sectors, can be described by the matrix fractional differential equation

$$(\widehat{D_{0+}^\alpha} Y)(t) = \Lambda \cdot Y(t), \qquad (51)$$

where $\Lambda = (E - A) \cdot B^{-1}$. The solution of equation (51) cannot be represented in the form $Y(t) = Y \cdot E_\alpha[\lambda \cdot t^\alpha]$. It can be found in the form

$$Y(t) = \widehat{E_\alpha}[\lambda \cdot t^\alpha] \cdot Y, \qquad (52)$$

where $\widehat{E_\alpha}[\lambda \cdot t^\alpha] = \text{diag}(E_{\alpha_1}[\lambda \cdot t^{\alpha_1}], \ldots, E_{\alpha_n}[\lambda \cdot t^{\alpha_n}])$ is the square diagonal matrix of the Mittag-Leffler functions (21). For the two-sector model the matrix function $\widehat{E_\alpha}[\lambda \cdot t^\alpha]$ can be written in the form

$$\widehat{E_\alpha}[\lambda \cdot t^\alpha] = \begin{pmatrix} E_{\alpha_1}[\lambda \cdot t^{\alpha_1}] & 0 \\ 0 & E_{\alpha_2}[\lambda \cdot t^{\alpha_2}] \end{pmatrix}. (53)$$

It is easy to check the property of the matrix function $\widehat{E_\alpha}[\lambda \cdot t^\alpha]$, that is given by the equation

$$\widehat{D_{0+}^\alpha} \widehat{E_\alpha}[\lambda \cdot t^\alpha] = \lambda \cdot \widehat{E_\alpha}[\lambda \cdot t^\alpha]. \qquad (54)$$

Using (54), the vectors $Y$ must satisfy the equation

$$(\Lambda - \lambda \cdot E) \cdot Y = 0. \qquad (55)$$

In order to equation (55) holds for the non-trivial vector $Y$, it is necessary and sufficient that the matrix $(\Lambda - \lambda \cdot E)$ is singular, i.e., its determinant should be equal to zero: $|\Lambda - \lambda \cdot E| = 0$. For the roots $\lambda_k$ of the characteristic equation $|\Lambda - \lambda \cdot E| = 0$, we can find the nonzero vectors $Y_k = (Y_{kj})$ by using equation (55). As a result, the general solution of fractional differential equations (51) with $0 < \alpha_k < 1$ for all k=1,…,n can be written in the form

$$Y(t) = \sum_{k=1}^n c_k \cdot \widehat{E_\alpha}[\lambda_k \cdot t^{\alpha_k}] \cdot Y_k. \qquad (56)$$

In the component form, equation (56) can be written as

$$Y_j(t) = \sum_{k=1}^n c_k \cdot E_{\alpha_j}[\lambda_k \cdot t^{\alpha_j}] \cdot Y_{kj}, \qquad (57)$$

where $\lambda_k$ are eigenvalues of the matrix $\Lambda := (E - A) \cdot B^{-1}$, and $Y_k = (Y_{kj})$ are the corresponding eigenvectors ($\Lambda \cdot Y_k = \lambda_k \cdot Y_k$). The coefficients $c_k$ are determined by the initial conditions

$$\sum_{k=1}^n c_k \cdot Y_k = Y(0). \qquad (58)$$

Equation (56) at t = 0 gives (58) since $\widehat{E_\alpha}[0] = E$, where E is the unit diagonal matrix n-th order.

Using (29) for real values of $\lambda$, we find that the growth rates are determined by the values

$$\lambda_{k,\text{eff}}(\alpha_k) := \lambda_k^{1/\alpha_k}, \qquad (59)$$

which are interpreted as the effective growth rate of the k-th sector. In contrast to the case (32), the effective growth rates are changes not equally over the standard model. As a result, in the ordered series of the growth rates $\lambda_k$, the order can be changed, when we take into account the memory effects.

In solution (56) at $t \to \infty$ the term with the maximum real part $\lambda_{k,\text{eff}}(\alpha_k)$, for which $c_k \neq 0$ and $\theta$ satisfies inequality (30). If the dominant term has the effective growth rate

$$\lambda_{s,\text{eff}}(\alpha_s) = (1/s_{\max})^{1/\alpha_s}, \qquad (60)$$

where $s_{\max}$ is the Frobenius-Perron number of the matrix S, then the growth rate of the final product (national income) for all sectors will tend to the effective technological growth rate (60) at $t \to \infty$. The sectoral structure of the national income in the limit will be determined by the proportions between the coordinates of the eigenvector $Y_s$ of matrix $\Lambda$, which corresponds to $\lambda_s = 1/s_{\max}$.

Note that the effective growth rate may not be the same in different sectors and different eigenvalues. The condition of equality of the effective growth rates can be expressed by the equation

$$\frac{\alpha_k}{\alpha_l} = \frac{\ln(\lambda_k)}{\ln(\lambda_l)}. \qquad (61)$$

For the case $1 < \alpha_k < 2$ for all k=1,…,n, the general solution of the fractional differential equation

$$\left(\widehat{D_{0+}^\alpha} Y\right)(t) = \Lambda \cdot Y(t) \qquad (62)$$

with the matrix Caputo fractional derivative (48) of order 1<α<2 can be written [12, p. 232] as

$$Y(t) = \sum_{k=1}^n c_{1k} \cdot \widehat{E_\alpha}[\lambda_k \cdot t^\alpha] \cdot Y_k + \sum_{k=1}^n c_{2k} \cdot t \cdot \widehat{E_{\alpha,2}}[\lambda_k \cdot t^\alpha] \cdot Y_k, \qquad (63)$$

where $\widehat{E_{\alpha,2}}[\lambda \cdot t^\alpha] = \text{diag}(E_{\alpha_1,2}[\lambda \cdot t^{\alpha_1}], \dots, E_{\alpha_n,2}[\lambda \cdot t^{\alpha_n}])$ is the square diagonal matrix of the two-parametric Mittag-Leffler functions (36).

If a part of the memory fading parameters belongs to the interval $\alpha_i \in (0,1)$, and the other part of the parameters is in the range $\alpha_j \in (1,2)$, then in the solution (63) will be absent terms with $\widehat{E_{\alpha,2}}[\lambda_k \cdot t^\alpha]$ for $\alpha_i \in (0,1)$.

The open dynamic intersectoral model with sectoral memory and non-zero non-productive consumption (C(t) ≠ 0) is described by the matrix fractional differential equation

$$\left(\widehat{D_{0+}^\alpha} Y\right)(t) = (E - A) \cdot B^{-1} \cdot Y(t) - (E - A) \cdot B^{-1} \cdot C(t). \qquad (64)$$

The general solution of fractional differential equations (64) can be represented as the sum of the general solution $Y_0(t)$ of the closed model, which is described by homogeneous equation (56), and a particular solution $Y_C(t)$ of the inhomogeneous equation (64), such that

$$Y(t) = Y_0(t) + Y_C(t). \qquad (65)$$

Let us consider the vector function of non-productive consumption C(t) in the form

$$C(t) = \widehat{E_\alpha}[r \cdot t^\alpha] \cdot C(0), \qquad (66)$$

where $\widehat{E_\alpha}[r \cdot t^\alpha]$ is the square diagonal matrix, and C(0) is the vector, which defines non-productive consumption in sectors at the initial time t=0. Here we do not impose the restriction that components of the vector C(t) has the same constant growth rate. The components of the vector C(t) has the form $C_k(t) = E_{\alpha_k}[r_k \cdot t^{\alpha_k}] \cdot C_k(0)$, where $r_k$ is the growth rate of non-productive consumption in the k-th sector.

As a result, the particular solution of equation (64) is given by the expression

$$Y_C(t) = (E - R \cdot S)^{-1} \cdot \widehat{E_\alpha}[r \cdot t^\alpha] \cdot C(0), \qquad (67)$$

where $S = B \cdot (E - A)^{-1}$ is the matrix of the full incremental capital intensity, and $R = \text{diag}(r_1, \dots, r_n)$ is the matrix of the growth rate of non-productive consumption. If $0 < \alpha_k < 1$ for all k=1,…,n, the general solution of the open dynamic intersectoral model with sectoral memory can be written in the form

$$Y(t) = \sum_{k=1}^n c_k \cdot \widehat{E_\alpha}[\lambda_k \cdot t^\alpha] \cdot Y_k + (E - R \cdot S)^{-1} \cdot \widehat{E_\alpha}[r \cdot t^\alpha] \cdot C(0), \qquad (68)$$

where $\lambda_k$ are eigenvalues of the matrix $\Lambda \coloneqq (E - A) \cdot B^{-1}$, and $Y_k = (Y_{kj})$ are the corresponding eigenvectors ($\Lambda \cdot Y_k = \lambda_k \cdot Y_k$). Thee coefficients $c_k$ are determined (for the case $0 < \alpha_k < 1$ for all k=1,…,n) by the initial conditions

$$\sum_{k=1}^n c_k \cdot Y_k = Y(0) - (E - R \cdot S)^{-1} \cdot C(0). \qquad (69)$$

For R=r·E condition (69) takes the standard form (47).

For $1 < \alpha_k < 2$ for all k=1,…,n the solutions of open model with sectoral memory are obtained analogously the solutions from the closed model with $1 < \alpha_k < 2$ for all k=1,…,n.

A comparison of the growth rates of the intersectoral models with sectoral memory and the standard model without memory, is given in the following table for the real values of $\lambda_k$ and $0 < \alpha_k < 2$.

|  | $0 < \alpha_k < 1$ | $1 < \alpha_k < 2$ |
|---|---|---|
| $0 < \lambda_k < 1$ | $\lambda_{k,eff}(\alpha_k) < \lambda_k$ | $\lambda_{k,eff}(\alpha_k) > \lambda_k$ |
| $\lambda_k = 1$ | $\lambda_{k,eff}(\alpha_k) = \lambda_k$ | $\lambda_{k,eff}(\alpha_k) = \lambda_k$ |
| $\lambda_k > 1$ | $\lambda_{k,eff}(\alpha_k) > \lambda_k$ | $\lambda_{k,eff}(\alpha_k) < \lambda_k$ |

The table shows that the accounting of sectoral memory can significantly change the growth rates of the economy and its sectors in the framework of the dynamic intersectoral models. At the same time the growth rates can increase and decrease in comparison with the standard intersectoral model without memory.

Using these results, we can formulate the following principle.

**Principle of domination change.** *In intersectoral economic dynamics the effects of fading sectoral memory can change the dominating behavior of economic sectors.*

For example, according to this principle, the inequality $\lambda_1 < \lambda_2$, of the standard model can leads to the inequality $\lambda_{1,eff}(\alpha_1) > \lambda_{2,eff}(\alpha_2)$ of the model with power-law sectoral memory. More detailed example that illustrates this principle is presented in the next section (see example 3).

## 6. EXAMPLES OF INTERSECTORAL MODEL WITH MEMORY

For illustration, we present numerical examples of finding solutions of the closed dynamic intersectoral models with power-low memory. Solutions of the dynamic model of the final and gross products will be illustrated by the examples of two sectors of economy. For the calculations, we used the Maple computer algebra package.

**Example 1.** We define the matrices A and B in the form
$$A = \begin{pmatrix} 0.1 & 0.2 \\ 0.2 & 0.3 \end{pmatrix}; \quad B = \begin{pmatrix} 0.4 & 0.4 \\ 1.0 & 0.5 \end{pmatrix}. \tag{70}$$
The inverse matrix $B^{-1}$ and the matrix $(E - A)$ are represented by the expressions
$$B^{-1} = \begin{pmatrix} -2.5 & 2 \\ 5 & -2 \end{pmatrix}; \quad (E - A) = \begin{pmatrix} 0.9 & -0.2 \\ -0.2 & 0.7 \end{pmatrix}. \tag{71}$$
Then the products of these matrices have the form
$$\Lambda = (E - A) \cdot B^{-1} = \begin{pmatrix} -3.25 & 2.2 \\ 4.0 & -1.8 \end{pmatrix}, \tag{72}$$
$$\Omega = B^{-1} \cdot (E - A) = \begin{pmatrix} -2.65 & 1.9 \\ 4.9 & -2.4 \end{pmatrix}. \tag{73}$$
For matrices (72) and (73) equations (17) and (18) have the form
$$D_{0+}^\alpha \begin{pmatrix} Y_1(t) \\ Y_2(t) \end{pmatrix} = \begin{pmatrix} -3.25 & 2.2 \\ 4.0 & -1.8 \end{pmatrix} \cdot \begin{pmatrix} Y_1(t) \\ Y_2(t) \end{pmatrix}, \tag{74}$$
$$D_{0+}^\alpha \begin{pmatrix} X_1(t) \\ X_2(t) \end{pmatrix} = \begin{pmatrix} -2.65 & 1.9 \\ 4.9 & -2.4 \end{pmatrix} \cdot \begin{pmatrix} X_1(t) \\ X_2(t) \end{pmatrix}, \tag{75}$$
where $0 < \alpha < 1$. The characteristic equations for matrices (72) and (73) have the form
$$\begin{vmatrix} -3.25 - \lambda & 2.2 \\ 4.0 & -1.8 - \lambda \end{vmatrix} = 0, \tag{76}$$
$$\begin{vmatrix} -2.65 - \omega & 1.9 \\ 4.9 & -2.4 - \omega \end{vmatrix} = 0, \tag{77}$$
which lead to the quadratic equations
$$\lambda^2 + 5.05 \cdot \lambda - 2{,}95 = 0. \tag{78}$$

$$\omega^2 + 5.05 \cdot \omega - 2{,}95 = 0. \tag{79}$$

The roots of characteristic equations (78) and (79) are the numbers

$$\lambda_1 \approx 0.5287886305; \quad \lambda_2 \approx -5.578788631, \tag{80}$$

where $\omega_1 = \lambda_1$, $\omega_2 = \lambda_2$. As a result, we have equations for the eigenvectors

$$\begin{pmatrix} -3.25 & 2.2 \\ 4.0 & -1.8 \end{pmatrix} \cdot Y_k = \lambda_k \cdot Y_k, \tag{81}$$

$$\begin{pmatrix} -2.65 & 1.9 \\ 4.9 & -2.4 \end{pmatrix} \cdot X_k = \omega_k \cdot X_k, \tag{82}$$

where k=1, 2, and $\lambda_k = \omega_k$ take values (80). Solutions of equations (81) and (82) gives the eigenvectors

$$Y_1 = \begin{pmatrix} 0.5245339728 \\ 0.9009559150 \end{pmatrix}; \quad Y_2 = \begin{pmatrix} 0.6867206226 \\ -0.7269214446 \end{pmatrix}. \tag{83}$$

$$X_1 = \begin{pmatrix} 0.5716020012 \\ 0.9563168119 \end{pmatrix}; \quad X_2 = \begin{pmatrix} 0.5442405426 \\ -0.8389292175 \end{pmatrix}. \tag{84}$$

Coordinates of eigenvectors (83) and (84) were obtained by using the Maple computer algebra package. Coefficients $c_k$ can be derived from the initial conditions by using equation (24) in the form

$$c_1 \cdot Y_1 + c_2 \cdot Y_2 = Y(0). \tag{85}$$

Let us use the initial conditions for the final product in the form

$$Y(0) = \begin{pmatrix} Y_1(0) \\ Y_2(0) \end{pmatrix} = \begin{pmatrix} 40 \\ 40 \end{pmatrix}. \tag{86}$$

Then the system of linear equations (85) with vectors (83) has the form

$$\begin{cases} 0.5245339728 \cdot c_1 + 0.6867206226 \cdot c_2 = 40, \\ 0.9009559150 \cdot c_1 - 0.7269214446 \cdot c_2 = 40. \end{cases} \tag{87}$$

The system of algebraic equations (87) have the solution $c_1 \approx 15.05687769$ and $c_2 \approx 56.54568268$. As a result, we get the functions of the final product of the first and second sectors

$$\begin{cases} Y_1(t) = 29.66013158 \cdot E_\alpha[0.5287886305 \cdot t^\alpha] + 10.33986842 \cdot E_\alpha[-5.578788631 \cdot t^\alpha], \\ Y_2(t) = 50.94516728 \cdot E_\alpha[0.5287886305 \cdot t^\alpha] - 10.94516728 \cdot E_\alpha[-5.578788631 \cdot t^\alpha]. \end{cases} \tag{88}$$

Using $E_\alpha[0] = 1$, we can see that solutions (88) at the initial time (t=0) gives the final product of the first and second sectors

$Y_1(0) = 29.66013158 + 10.33986842 = 40$,

$Y_2(0) = 50.94516728 - 10.94516728 = 40$,

which coincides with the initial conditions (86). For α=1, solution (88) takes the form

$$\begin{cases} Y_1(t) = 29.66013158 \cdot \exp(0.5287886305 \cdot t) + 10.33986842 \cdot \exp(-5.578788631 \cdot t), \\ Y_2(t) = 50.94516728 \cdot \exp(0.5287886305 \cdot t) - 10.94516728 \cdot \exp(-5.578788631 \cdot t). \end{cases} \tag{89}$$

where we use the property $E_1[z] = \exp(z)$.

For the vector X(t) of gross product, the coefficients $d_k$ are determined by the initial conditions

$$d_1 \cdot X_1 + d_2 \cdot X_2 = X(0), \tag{90}$$

where $X_1$ and $X_2$ are defined by expressions (84). Let us consider the initial conditions for the gross product in the form

$$X(0) = \begin{pmatrix} X_1(0) \\ X_2(0) \end{pmatrix} = \begin{pmatrix} 50 \\ 30 \end{pmatrix}. \tag{91}$$

Then the system of equations (90) with the vectors (84) and (91) has the form

$$\begin{cases} 0.5716020012 \cdot d_1 + 0.5442405426 \cdot d_2 = 50, \\ 0.9563168119 \cdot d_1 - 0.8389292175 \cdot d_2 = 30. \end{cases} \tag{92}$$

Solving this system of linear equations, we get $d_1 \approx 58.27367714$ and $d_2 \approx 30.66778055$.

Using the Maple to obtain the functions for gross product of the first and second sectors, we get

$$\begin{cases} X_1(t) = 33.30935047 \cdot E_\alpha[0.5287886305 \cdot t^\alpha] + 16.69064953 \cdot E_\alpha[-5.578788631 \cdot t^\alpha], \\ X_2(t) = 55.72809714 \cdot E_\alpha[0.5287886305 \cdot t^\alpha] - 25.72809714 \cdot E_\alpha[-5.578788631 \cdot t^\alpha]. \end{cases} \quad (93)$$

For α=1, solution (93) is written as the linear combination of exponents

$$\begin{cases} X_1(t) = 33.30935047 \cdot \exp(0.5287886305 \cdot t) + 16.69064953 \cdot \exp(-5.578788631 \cdot t), \\ X_2(t) = 55.72809714 \cdot \exp(0.5287886305 \cdot t) - 25.72809714 \cdot \exp(-5.578788631 \cdot t). \end{cases} \quad (94)$$

Let us find the Frobenius-Perron number of the matrix $S = B \cdot (E - A)^{-1} = \Lambda^{-1}$ of the full incremental capital intensity. Using matrices (70), we get

$$S = B \cdot (E - A)^{-1} = \begin{pmatrix} 0.6101694915 & 0.7457627119 \\ 1.355932203 & 1.101694915 \end{pmatrix}. \quad (95)$$

We see that the matrix S is positive. Using the Maple package, we obtain the eigenvalues of the matrix (95) in the form

$$s_1 \approx 1.891114790; \quad s_2 \approx -0.179250383. \quad (96)$$

As a result, the Frobenius-Perron number is equal to $s_{max} = s_1 \approx 1.891114790$. The corresponding eigenvalue $\Lambda = S^{-1}$, which describes the technological growth rate of the standard model without memory (with α=1), has the value

$$\lambda_s = \frac{1}{s_{max}} \approx 0.5287886305. \quad (97)$$

We can see that $\lambda_s = \lambda_1 \approx 0.5287886305$. In the closed intersectoral model (74), (75) with power-law memory, the growth rate is determined by the effective technological growth rate $\lambda_{s,eff}(\alpha) = (1/s_{max})^{1/\alpha} = (0.5287886305)^{1/\alpha}$. Eigenvalue (97) corresponds to the eigenvector $Y_1 = \begin{pmatrix} 0.5245339728 \\ 0.9009559150 \end{pmatrix}$, which determines the structure of the vector Y(t).

It should be noted that the second terms of expressions (89) and (93) contain the function $E_\alpha[-5.578788631 \cdot t^\alpha]$, the value of which decreases faster than the value of the first term $E_\alpha[0.5287886305 \cdot t^\alpha]$ increases. Therefore, the second terms of (89) and (93) can be neglected for long intervals of time. In this case, the final product (national income) of the economic sectors will be described by the equations

$$\begin{cases} Y_1(t) \approx 29.66013158 \cdot E_\alpha[0.5287886305 \cdot t^\alpha], \\ Y_2(t) \approx 50.94516728 \cdot E_\alpha[0.5287886305 \cdot t^\alpha], \end{cases} \quad (98)$$

and the gross product in these sectors will be described by the equations

$$\begin{cases} X_1(t) \approx 33.30935047 \cdot E_\alpha[0.5287886305 \cdot t^\alpha], \\ X_2(t) \approx 55.72809714 \cdot E_\alpha[0.5287886305 \cdot t^\alpha]. \end{cases} \quad (99)$$

As a result, the growth rate of national income and its sector structure quickly approaching $\lambda_s = \lambda_1$ and $Y_1$, respectively. The growth rate of gross product and sector structure are fast approaching to $\omega_s = \omega_1 = \lambda_1 = \lambda_s$ and $X_1$, respectively.

As a result, for this example we can conclude the following.

(a) The process of the national income of each sector is described by two terms with the Mittag-Leffler functions [25];

(b) The second terms of functions of the gross and final products for both sectors are described by rapidly decreasing processes and therefore can be ignored on long time intervals;

(c) The unit value of fading parameter corresponds to the standard model with absence of memory.

**Example 2.** Let us consider the matrices A and B of the following form

$$A = \begin{pmatrix} 0.1 & 0.2 \\ 0.2 & 0.3 \end{pmatrix}; \quad B = \begin{pmatrix} 0.4 & 0.2 \\ 1.0 & 0.9 \end{pmatrix}. \tag{100}$$

The matrix A is taken with the same values of the coefficients, as in Example 1. The matrix $(E - A)$ and the inverse matrix $B^{-1}$ are given by the expressions

$$(E - A) = \begin{pmatrix} 0.9 & -0.2 \\ -0.2 & 0.7 \end{pmatrix}; \quad B^{-1} = \begin{pmatrix} 5.625 & -1.25 \\ -6.25 & 2.5 \end{pmatrix}. \tag{101}$$

Then the product of these matrices has the form

$$\Lambda = (E - A) \cdot B^{-1} = \begin{pmatrix} 6.3125 & -1.625 \\ -5.5 & 2.0 \end{pmatrix}. \tag{102}$$

Equation (18) with matrix (102) has the form

$$D_{0+}^{\alpha} \begin{pmatrix} Y_1(t) \\ Y_2(t) \end{pmatrix} = \begin{pmatrix} 6.3125 & -1.625 \\ -5.5 & 2.0 \end{pmatrix} \cdot \begin{pmatrix} Y_1(t) \\ Y_2(t) \end{pmatrix}, \tag{103}$$

where we will assume $0<\alpha<1$. The characteristic equation for matrix (102) is written as

$$\begin{vmatrix} 6.3125 - \lambda & -1.625 \\ -5.5 & 2.0 - \lambda \end{vmatrix} = 0. \tag{104}$$

The roots of characteristic equation (104) are the numbers

$$\lambda_1 \approx 0.470206855; \quad \lambda_2 \approx 7.842293144. \tag{105}$$

The corresponding equations for the eigenvectors $Y_k$ have the form

$$\begin{pmatrix} 6.3125 & -1.625 \\ -5.5 & 2.0 \end{pmatrix} \cdot Y_k = \lambda_k \cdot Y_k, \tag{106}$$

where k=1, 2, and the values of $\lambda_k$ are given by (105). Solutions of equation (106) are the eigenvectors

$$Y_1 = \begin{pmatrix} 0.3027353638 \\ 1.088411532 \end{pmatrix}; \quad Y_2 = \begin{pmatrix} 0.7281147014 \\ -0.6854553099 \end{pmatrix}. \tag{107}$$

Coordinates of eigenvectors (107) were obtained by using the Maple package. Coefficients $c_k$ can be derived from the initial conditions (86) in the form

$$\begin{cases} 0.3027353638 \cdot c_1 + 0.7281147014 \cdot c_2 = 40, \\ 1.088411532 \cdot c_1 - 0.6854553099 \cdot c_2 = 40. \end{cases} \tag{108}$$

The system (110) have the solution $c_1 \approx 56.54280044$ and $c_2 \approx 31.42704672$.

As a result, the functions of the final product (national income) of the first and second sectors are given by the expressions

$$\begin{cases} Y_1(t) = 17.11750526 \cdot E_\alpha[0.470206855 \cdot t^\alpha] + 22.88249474 \cdot E_\alpha[7.842293144 \cdot t^\alpha], \\ Y_2(t) = 61.54183605 \cdot E_\alpha[0.470206855 \cdot t^\alpha] - 21.54183605 \cdot E_\alpha[7.842293144 \cdot t^\alpha]. \end{cases} \tag{109}$$

Comparing (109) with (88) of Example 1, we can see that the first term is changed a little. In particular,

$$\lambda_s = \frac{1}{s_{max}} \approx 0.470206855. \tag{110}$$

At the same time, the second term has become dominant. Effective technological growth rate is $\lambda_{2,eff}(\alpha) := \lambda_2^{1/\alpha} = (7.842293144)^{1/\alpha}$. The second component of the vector Y(t) quickly reaches zero and then rapidly decreases. As a result, we can conclude that the solution of the closed model with matrix (100), gives unacceptable results, both for the standard model without memory and for the model with power-law memory.

**Example 3.** Let us consider intersectoral model with two sectors, in which the memory fading parameters of these sectors are not the same. We consider the matrices

$$A = \begin{pmatrix} 0.1 & 0.2 \\ 0.2 & 0.3 \end{pmatrix}; \quad B = \begin{pmatrix} 0.3 & 0.1 \\ 0.2 & 0.3 \end{pmatrix}. \tag{111}$$

where the matrix A is taken with the same coefficients, as in Example 1. The matrices $(E - A)$ and $B^{-1}$ are given by the expressions

$$(E - A) = \begin{pmatrix} 0.9 & -0.2 \\ -0.2 & 0.7 \end{pmatrix}, \tag{112}$$

$$B^{-1} = \begin{pmatrix} 4.285714286 & -1.428571429 \\ -2.857142857 & 4.285714286 \end{pmatrix}. \tag{113}$$

Then the product of matrices (112) and (113) has the form

$$\Lambda = (E - A) \cdot B^{-1} = \begin{pmatrix} 4.428571428 & -2.142857143 \\ -2.857142857 & 3.285714286 \end{pmatrix}. \tag{114}$$

Equation (18) with matrix (102) has the form

$$\widehat{D_{0+}^{\alpha}} \begin{pmatrix} Y_1(t) \\ Y_2(t) \end{pmatrix} = \begin{pmatrix} 4.428571428 & -2.142857143 \\ -2.857142857 & 3.285714286 \end{pmatrix} \cdot \begin{pmatrix} Y_1(t) \\ Y_2(t) \end{pmatrix}, \tag{115}$$

where $0 < \alpha_1 < 1$, $0 < \alpha_2 < 1$ and $\alpha_1 \neq \alpha_2$ in general. The characteristic equation for matrix (114) is written as

$$\begin{vmatrix} 4.428571428 - \lambda & -2.142857143 \\ -2.857142857 & 3.285714286 - \lambda \end{vmatrix} = 0. \tag{116}$$

The roots of characteristic equation (116) are the numbers

$$\lambda_1 \approx 1.317658738; \; \lambda_2 \approx 6.396626976. \tag{117}$$

The corresponding equations for the eigenvectors $Y_k$ have the form

$$\begin{pmatrix} 4.428571428 & -2.142857143 \\ -2.857142857 & 3.285714286 \end{pmatrix} \cdot Y_k = \lambda_k \cdot Y_k, \tag{118}$$

where $\lambda_k$ is defined by values (117) and k=1, 2. Solutions of equation (118) are the eigenvectors

$$Y_1 = \begin{pmatrix} 0.5728488169 \\ 0.8316385716 \end{pmatrix}; \; Y_2 = \begin{pmatrix} 0.7365083952, \\ -0.6764284024 \end{pmatrix}. \tag{119}$$

Coefficients $c_k$ are determined by the initial

$$c_1 \cdot Y_1 + c_2 \cdot Y_2 = Y(0). \tag{120}$$

We consider the initial conditions for the vector of the final product in the form

$$Y(0) = \begin{pmatrix} Y_1(0) \\ Y_2(0) \end{pmatrix} = \begin{pmatrix} 40 \\ 40 \end{pmatrix}. \tag{121}$$

Then the system of equations (120) with vectors (119) has the form

$$\begin{cases} 0.5728488169 \cdot c_1 + 0.7365083952 \cdot c_2 = 40, \\ 0.8316385716 \cdot c_1 - 0.6764284024 \cdot c_2 = 40. \end{cases} \tag{122}$$

Solving the system of equations (122), we obtain the values $c_1 \approx 56.51747192$ and $c_2 \approx 10.35159019$. As a result, the final product (national income) of the first and second sectors are given by the expressions

$$\begin{cases} Y_1(t) = 32.37596692 \cdot E_{\alpha_1}[1.317658738 \cdot t^{\alpha_1}] + 7.624033079 \cdot E_{\alpha_2}[6.396626976 \cdot t^{\alpha_2}], \\ Y_2(t) = 47.00210962 \cdot E_{\alpha_1}[1.317658738 \cdot t^{\alpha_1}] - 7.002109615 \cdot E_{\alpha_2}[6.396626976 \cdot t^{\alpha_2}]. \end{cases} \tag{123}$$

For $\alpha_1 = \alpha_2 = 1$, equation (123) describes the standard model, since $E_1[z] = \exp(z)$.

Let us find the Frobenius-Perron number of the matrix $S = B \cdot (E - A)^{-1} = \Lambda^{-1}$ of the full incremental capital intensity. Using (111), we get

$$S = B \cdot (E - A)^{-1} = \begin{pmatrix} 0.3898305085 & 0.2542372881 \\ 0.3389830508 & 0.5254237289 \end{pmatrix}. \tag{124}$$

Using the Maple software package, we obtain the eigenvalues of this matrix

$$s_1 \approx 0.1563323927; \; s_2 \approx 0.7589218447. \tag{125}$$

As a result, the Frobenius-Perron number is equal to $s_{max} = s_2 \approx 0.7589218447$. The corresponding eigenvalue $\Lambda = S^{-1}$, which describes the technological growth rate of the standard model without memory (with α=1) has the form

$$\lambda_s = \frac{1}{s_{max}} \approx 1.317658738. \tag{126}$$

Eigenvalue (126) corresponds to the eigenvector $Y_1 = \begin{pmatrix} 0.5728488169 \\ 0.8316385716 \end{pmatrix}$, which determines the structure of the vector Y(t).

In the intersectoral model with power-law memory, the growth rate is determined by the effective technological growth rate (60), which has the form

$$\lambda_{s,\text{eff}}(\alpha_1) := \lambda_1^{1/\alpha_1} = (1.317658738)^{1/\alpha_1}. \tag{127}$$

In the same time, for the second sector the effective growth rate (59) is equal to

$$\lambda_{2,\text{eff}}(\alpha_2) := \lambda_2^{1/\alpha_2} = (6.396626976)^{1/\alpha_2}. \tag{128}$$

For $\alpha_1 = 0.1$ and $\alpha_2 = 0.9$, we get the values

$$\lambda_{s,\text{eff}}(\alpha_1) := \lambda_1^{1/\alpha_1} = 15.77718303, \tag{129}$$

$$\lambda_{2,\text{eff}}(\alpha_2) := \lambda_2^{1/\alpha_2} = 7.861422531. \tag{130}$$

As a result, for this example we can conclude the following.

(a) For standard technological growth rates, we have the inequality $\lambda_s = \lambda_1 < \lambda_2$. As a result, the standard intersectoral model without memory gives unacceptable results. This statement is caused by the dominance of the trajectory $\exp(6.396626976 \cdot t)$ at $t \to \infty$, which leads to the fact that the second component of the vector Y(t) quickly reaches zero and then rapidly decreases.

(b) For intersectoral model with sectoral memory, we have the reverse inequality $\lambda_{s,\text{eff}}(\alpha_2) = \lambda_{1,\text{eff}}(\alpha_1) > \lambda_{2,\text{eff}}(\alpha_2)$. This intersectoral model with sectoral power-law memory, which based on the same matrices A and B as in the standard model, gives acceptable results with economic sense. This statement is based on the dominance of the trajectory $E_\alpha[1.317658738 \cdot t^\alpha]$ at $t \to \infty$, for which both components of the vector Y(t) increases with the rate of (129).

(c) As a result, it is clear that the inclusion of sectoral memory into intersectoral model with matrices (111), may lead to changes of dominance. This means that memory effects can lead to qualitatively different results compared to the standard model. In accordance with the proposed principle of domination change, the effects of sectoral memory can change the dominating behavior of economic sectors.

## 7. CONCLUSION

Memory effects can play an important role in finance [35, 36, 37, 38, 39, 40, 41, 42, 43, 44, 45, 46, 47, 48] and economics [27, 28, 29, 30, 31, 32, 33, 34]. Neglect of the memory effects in economic models, may lead to a distortion of the results in macroeconomic models [27, 28, 29, 30, 31, 32, 33, 34], and in intersectoral dynamic models. Therefore researches of the economic growth and the construction of appropriate economic models should take into account the possible dependence of economic processes on memory effects. The approach, which is proposed in this paper, can be considered as a theoretical and methodological basis for the application of dynamic intersectoral models with matrices of input-output balance to describe economic processes with power-law memory.

In this article, we proved that the inclusion of memory effects into the dynamic intersectoral models can lead to qualitatively new results at the same the matrices of the direct material costs, the incremental capital intensity of production and the full incremental capital intensity. The proposed approach to economic dynamics allows us to build more adequate intersectoral dynamic models of the economy. This approach takes into account that economic agents may remember the story of changes in economic processes (endogenous and exogenous variables) and the agents can take into account these changes in making economic decisions. In addition, we think that the parameters of the memory fading can be considered as control parameters to increase economic growth in the economic policy and the impacts on the real economy.

**Appendix: Dynamic memory and fractional calculus**

In the standard dynamic Leontief model the relationship between the accumulation and of the growth of gross output is assumed, that is, the model disregards time delay and memory effects. To take into account the memory effects, we can describe dependence of the vector I(t) of capital investments on the vector X(t) of the gross product by the equations

$$I(t) = B \cdot \int_0^t M_X(t-\tau) \cdot X^{(n)}(\tau) d\tau, \tag{A1}$$

$$\int_0^t M_I(t-\tau) \cdot I(\tau) d\tau = B \cdot X^{(n)}(t), \tag{A2}$$

where $M_X(t-\tau)$ and $M_I(t-\tau)$ are function, which are called the memory functions (or the linear response functions), and $X^{(n)}(\tau)$ is the derivative of the integer order n≥0. Equation (A1) allows us to describe the time delay and memory effects with respect to the gross product. Equation (A2) describes the delay and memory effects with respect to the capital investments.

For processes without dynamic memory, the memory functions are expressed in terms of the Dirac delta-function in the form $M_X(t-\tau) = M_I(t-\tau) = \delta(t-\tau)$, where $\delta(t-\tau)$ is the Dirac delta-function. The absence of memory means that the endogenous variable is determined by an exogenous variable X(t) only at the moment of time t. We can say that instantaneous forgetting of the history of the factor changes is realized. Substitution of $M_X(t-\tau) = M_I(t-\tau) = \delta(t-\tau)$, into equations (A1) and (A2) with n=1 gives equation (5) of economic accelerator $I(t) = B \cdot X^{(1)}(t)$ that describes process without dynamic memory and time delay. In this case, the process connects the sequence of subsequent states of the economic process to the previous state only through the current state for each time t. If the memory function M(t–τ) has the form M(t)=m·δ(t–T), then equations (A1) and (A2) can be written as an equation of accelerator with is a fixed-time delay of T periods, where the time-constant of the lag T is a given positive integer.

We can consider the memory function M(t–τ) that satisfies the normalization condition

$$\int_0^t M(\tau) d\tau = 1. \tag{A3}$$

In this case, the function M(t,τ)= M(t–τ) is often called the weighting function [49, p. 26]. The memory functions M(t), which satisfy the normalization condition, are often used to describe economic processes with continuously distributed lag [49, p. 25]. The existence of the time delay (lag) is connected with the fact that the processes take place with a finite speed, and the change of the economic factor does not lead to instant changes of indicator that depends on it. If normalization condition (A3) holds, the economic process goes through all the states without any losses. In this case we say that the memory function describes the complete (ideal) memory.

It should be noted that power laws play an important role in economics and finance [50, 51]. To describe the economic processes with power-law fading memory, we can use the memory function in the form

$$M_X(t-\tau) = \frac{1}{\Gamma(\beta)} \frac{m(\beta)}{(t-\tau)^{1-\beta}}, \tag{A4}$$

$$M_I(t-\tau) = \frac{1}{\Gamma(\gamma)} \frac{m(\gamma)}{(t-\tau)^{1-\gamma}}, \tag{A5}$$

where Γ(z) is the Gamma function, β and γ are parameters that characterize the power-law fading, t>τ. Here m(β) and m(γ) are positive real numbers with dimensions $t^{1-\beta}$ and $t^{1-\gamma}$, which are used to have easily interpretable dimension of the quantities. For simplification, we assume that these numbers are equal to one.

Substitution of memory functions (A4) into equations (A1) gives the fractional integral and differential equations of the order α>0 in the form

$$I(t) = B \cdot (D_{0+}^{\alpha} X)(t), \tag{A6}$$

where $\alpha = n - \beta > 0$, n:=[α]+1, and $D_{0+}^{\alpha}$ is the left-sided Caputo fractional derivative of the order α≥0 with respect to variable t>0 [12, p. 92]. This Caputo fractional derivative is defined by the equation

$$(D_{0+}^{\alpha} X)(t) := \frac{1}{\Gamma(n-\alpha)} \int_0^t \frac{X^{(n)}(\tau) d\tau}{(t-\tau)^{\alpha-n+1}}, \tag{A7}$$

where $\Gamma(z)$ is the Gamma function, t>0, and $X^{(n)}(\tau)$ is the derivative of the integer order n:=[α]+1 with respect to τ. It is assumed that the function X(τ) has derivatives up to the (n-1)th order, which are absolutely continuous functions on the interval [0,t].

Substitution of memory functions (A5) into equations (A2) gives the fractional integral and differential equations of the order α>0 in the form

$$(I_{0+}^{\gamma} I)(t) = B \cdot X^{(n)}(\tau), \tag{A8}$$

where $I_{0+}^{\gamma}$ is the left-sided Riemann-Liouville integral of the order γ>0 with respect to time variable. This integral is defined [12, p. 69-70] by the equation

$$(I_{0+}^{\gamma} I)(t) := \frac{1}{\Gamma(\gamma)} \int_0^t \frac{I(\tau) d\tau}{(t-\tau)^{1-\gamma}}, \tag{A9}$$

where the function I(t) is assumed to be measurable on the interval (0,T) and it must satisfy the condition $\int_0^t |I(\tau)| d\tau < \infty$. It is known that the Caputo derivative is inversed to the Riemann-Liouville integral [12, p. 96]. In other words, for any continuous function $I(\tau) \in C[0,t]$, the generalized Newton-Leibniz formula $(D_{0+}^{\gamma} I_{0+}^{\gamma} I)(t)$,, which is given by formula 2.4.32 in [12, p. 95], is satisfied. As a result, the action of the Caputo fractional derivative $D_{0+}^{\gamma}$ on equation (A8) gives

$$I(t) = B \cdot (D_{0+}^{\alpha} X)(t), \tag{A10}$$

where $\alpha = n + \gamma$, and $D_{0+}^{\alpha}$ is the left-sided Caputo fractional derivative (A7) of the order α≥0 with respect to variable t. We can see that equations (A6) and (A10) are the same up to interpretation of order α>0.

Equations (A6) and (A10) describe the same economic accelerator with the power-law memory of the order α≥0. As a result, we have a single equation that allows us to describe the power-law memory with respect to the gross product and the capital investments.